\documentclass[%
 aip,
 amsmath,amssymb,
 reprint,
]{revtex4-1}

\usepackage{graphicx}
\usepackage{dcolumn}
\usepackage{bm}

\usepackage[utf8]{inputenc}
\usepackage[T1]{fontenc}
\usepackage{mathptmx}
\usepackage{etoolbox}

\makeatletter
\def\@email#1#2{%
 \endgroup
 \patchcmd{\titleblock@produce}
  {\frontmatter@RRAPformat}
  {\frontmatter@RRAPformat{\produce@RRAP{*#1\href{mailto:#2}{#2}}}\frontmatter@RRAPformat}
  {}{}
}%
\makeatother
\begin{document}

\preprint{AIP/123-QED}

\title{Possible ferroic properties of copper-substituted lead phosphate apatite}

\author{Jiri Hlinka}
\affiliation{Institute of Physics$,$ Academy of Sciences of the Czech Republic$,$ Na Slovance 2$,$ 18221 Praha 8$,$ Czech Republic}\email{hlinka@fzu.cz}

\date{\today}

\begin{abstract}
Paper provides symmetry arguments explaining that charge density wave induced by copper doping to lead phosphate apatite crystal implies a symmetry breaking phase transition from $P6_3/m$ (176) to a polar and chiral phase with $P6_3$ (173) spacegroup symmetry.
 \end{abstract}

\maketitle

Recently reported claim for the room-temperature superconductivity in copper-doped lead phosphate apatite crystal denoted LK-99\cite{Lee233a, Lee236a} has already inspired several experimental\cite{Kumar23,Liu23} and theoretical\cite{Lai23, Griffin23, Si23, Wu23, Abramian23, Munoz23, Oh23} studies aimed to verify, clarify and understand these reported observations,  promising at least in principle a dramatic breakthrough in superconducting physics and technology.

The density functional theory confirmed the metal-insulator phase transition induced by copper doping to the parent insulating lead phosphate apatite.  At the level of the standard density functional theory, it was shown that the resulting copper-doped material Pb$_9$Cu(PO$_4$)$_6$O has a partially filled flat CuO conduction band\cite{Lai23, Griffin23, Si23, Munoz23}.
Substitution-induced structural changes resolved in calculations were similar to those inferred from the experiments, and included the overall unit cell volume shrinkage as well as local structural changes related to the lead ion lone pair charge density wave\cite{Griffin23}. Another theoretically confirmed salient and  potentially beneficial feature for the desired superconductivity transition is the remarkably strong electron-phonon coupling\cite{Munoz23}.

The powder diffraction experiments\cite{Lee236a,Lee233a} with Pb$_{9}$Cu(PO$_4$)$_6$O and the parent Pb$_{10}$(PO$_4$)$_6$O polycrystalline material proved hexagonal symmetry in both, about 0.5 percent shrinkage of the unit cell induced by copper doping and convincing signatures of modified in-plane lead-lead distances upon doping. In addition, it was inferred from these data that there is no structural phase transition\cite{Lee236a} induced by copper doping, meaning that the $P6_3$/m (176) spacegroup symmetry of the average crystal structure has not been altered through the metal-insulator phase transition. 
This conjecture can be extended also to the assumed superconducting phase transition, because the analysis surely involves the diffraction experiments performed at ambient temperature.  In fact, the $P6_3$/m (176)  structural variant is the most common one within the apatite family, and thus it is {\it a priori} the most likely option for the slightly doped material as well.

Nevertheless, 
 there is no single crystals diffraction experiment on Pb$_9$Cu(PO$_4$)$_6$)O available yet, so it is reasonable to cautiously  consider other plausible hexagonal structures, with similar local atomic arrangements, but belonging to other than the $P6_3$/m (176) spacegroup.  Moreover, some of these other variants are encountered in the  density functional calculations devoted to copper-doped lead phosphate apatite. The aim of this paper is to compare these plausible crystal structures,
and to address them in terms of macroscopic crystallography symmetry breaking, and their potential ferroic properties\cite{Aizu, Hlinka2016}. 
In particular, we argue that the reported structure factor analysis\cite{Lee233a} implies that the metal-insulator phase transition is also a ferroelectric transition, associated with the $P6_3$/m $>$ $P6_3$ symmetry-breaking of the average structure.

\begin{figure}[h]
\includegraphics[width=.7\columnwidth]{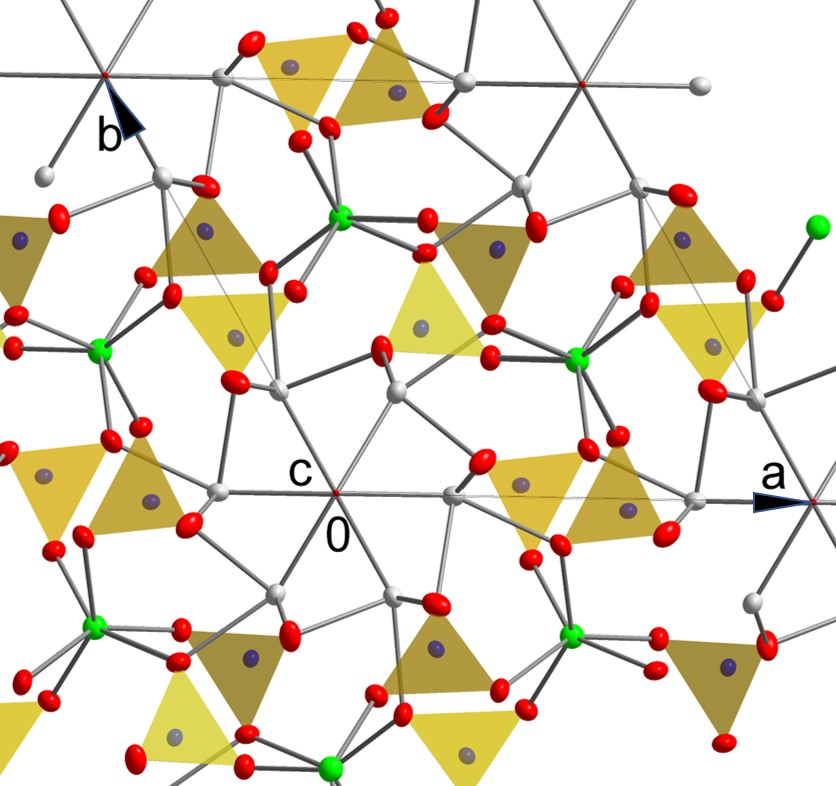}
\caption{The [001] orthographic projection of the crystal structure of the insulating parent compound Pb$_{10}$(PO$_4$)$_6$O according to the description of Ref.\,\onlinecite{Krivo03}. 
Color code: O - red, Pb(1) - gray, Pb(2) - green, P-blue, PO$_4$ - light to dark yellow tetrahedra\cite{Diamond,Fabry}. 
}
\label{prvni}
\end{figure}

\section{Parent $P6_3/m$ phase and $P6_3/mcm$ aristotype }

The $P6_3$/m (176) structure of the parent  Pb$_{10}$(PO$_4$)$_6$O apatite\cite{Krivo03} can be traditionally described in terms of the 
Pb$^{\rm F}_4$(PO$_4$)$_6$ framework with "tunnels" Pb$^{\rm T}_{6}$O, running along the hexagonal axes.
One of the "tunnel" axis passes through the origin of the conventional crystallographic unit cell labeled "0" in Fig.\,1. 
The inner tunnel radius can be defined as the distance from its axis to any of the nearest lead ions.
In other words, the inner wall of the tunnel can be associated with the  six  Pb$^{\rm T}$ lead ions located at Wyckoff site 6h
(grey Pb$^{\rm T}$ cations in Fig.\,1). The next nearest to the tunnel axis are oxygen anions of Wyckoff site 12i.
According to the structural refinement of Ref.\,\onlinecite{Krivo03}, the  "extra" O$^{\rm T}$ ion is disordered among 4 equivalent positions located on the axis of the tunnel in Wyckoff site 4e.

\begin{figure}[h]
\includegraphics[width=.95\columnwidth]{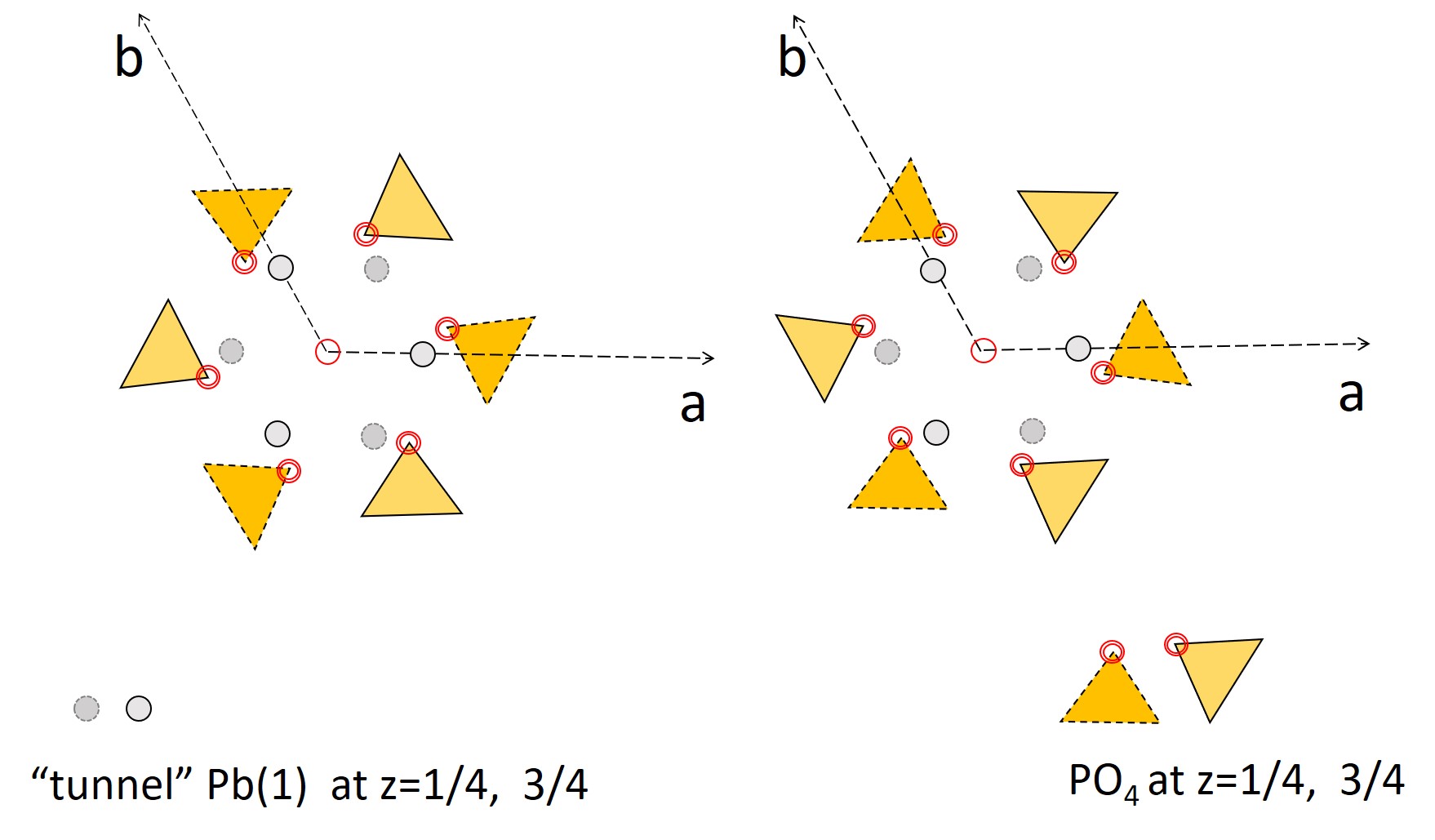}
\caption{Immediate vicinity of the structural tunnel in lead apatite structure. The color code for O$^{\rm T}$,  Pb$^{\rm T}$ and PO$_4$ tetrahedra with marked  12i oxygen anions is adopted from Fig.\,1.
}
\label{twins}
\end{figure}

For the sake of clarity, the immediate vicinity of the tunnel including  O$^{\rm T}$,  Pb$^{\rm T}$ and PO$_4$ tetrahedra with marked  12i oxygen anions is redrawn in Fig.\,2 in two different ways. On the left side, we show the original positions as given in Fig.\,1, and on the right side, we show the same structure, but flipped by the $m_{xz}$ mirror operation, passing through the {\bf a} and {\bf b} lattice vectors. We note that the $m_{xz}$ mirror is not a symmetry operation of the $6/m$ point group of the crystal structure but it is a symmetry operation of the $6/mmm$ point group of the underlying crystal lattice, meaning that the two atomic configurations displayed in Fig.\,2 represent realization of the two macroscopic merohedral twins guaranteed by the Friedel's law\cite{Boyer73}. Up to a fractional translation, the same result can be obtained by the other  11 symmetry operations lost in the $6/mmm  > 6/m$  symmetry reduction.

On the top of this purely symmetry property, it has been noticed that in the special case of apatite,  the twin generated by this particular twinning operation (and by the other  symmetry operations lost in the $P6_3/mcm$ (193) $>$ $P6_3/m$ (176) symmetry reduction) can be also generated from the original structure by relatively small atomic displacements only, comparable to  configuration changes in other materials with transformation twins.   
This subtle structural property allows to construct an auxiliary higher crystal structure,  $P6_3/mcm$ spacegroup symmetry, and then to consider the merohedral twins as the two transformation twins resulting from a hypothetical $P6_3/mcm > P6_3/m$ phase transitions\cite{Boyer73}.
There is certain liberty in the  definition of the exact atomic positions of the aristotype. 
However, for standard material-specific purposes, the most natural choice would be to derive the aristotype structure directly from the $P6_3/m$ structure. This can be conveniently done in the following four steps\cite{Boyer73}: (i) associate each atom with its closest counterpart in the $m_{xz}$-twin structure  (ii) define a new position of each cation as average between the original position and that of its partner, (iii) treat the PO$_4$ tetrahedron as a rigid body and achieve its symmetric orientation by a slight rotation around $z$ axis, (iv) the disordered tunnel oxygen remains at its original position.

\section{Possible lower-symmetry structures}

\begin{figure}[h]
\includegraphics[width=.95\columnwidth]{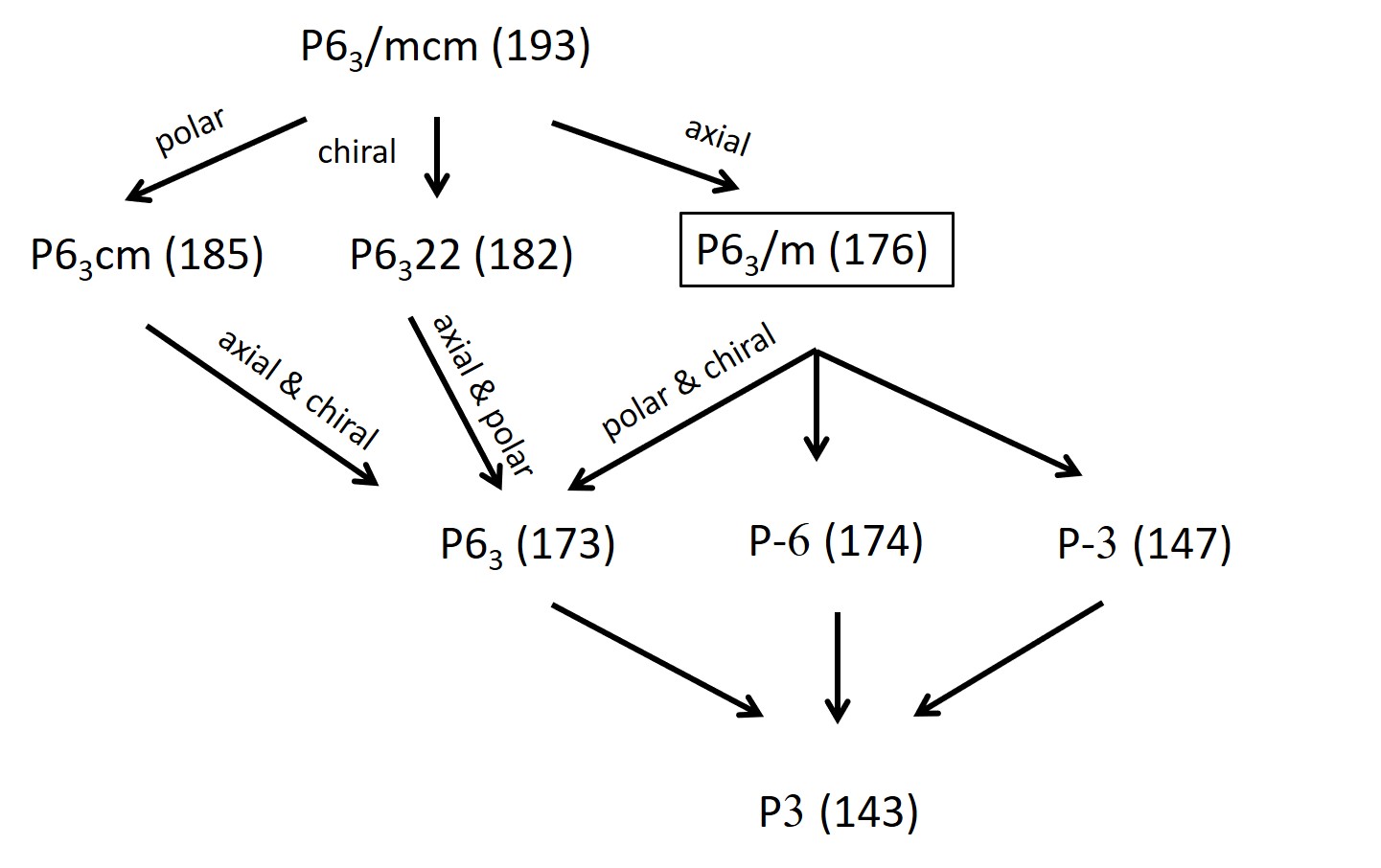}
\caption{ Subordination of hexagonal subgroups of $P6_3/mcm$ aristotype.
}
\label{druhy}
\end{figure}

\begin{figure}[h]
\includegraphics[width=1\columnwidth]{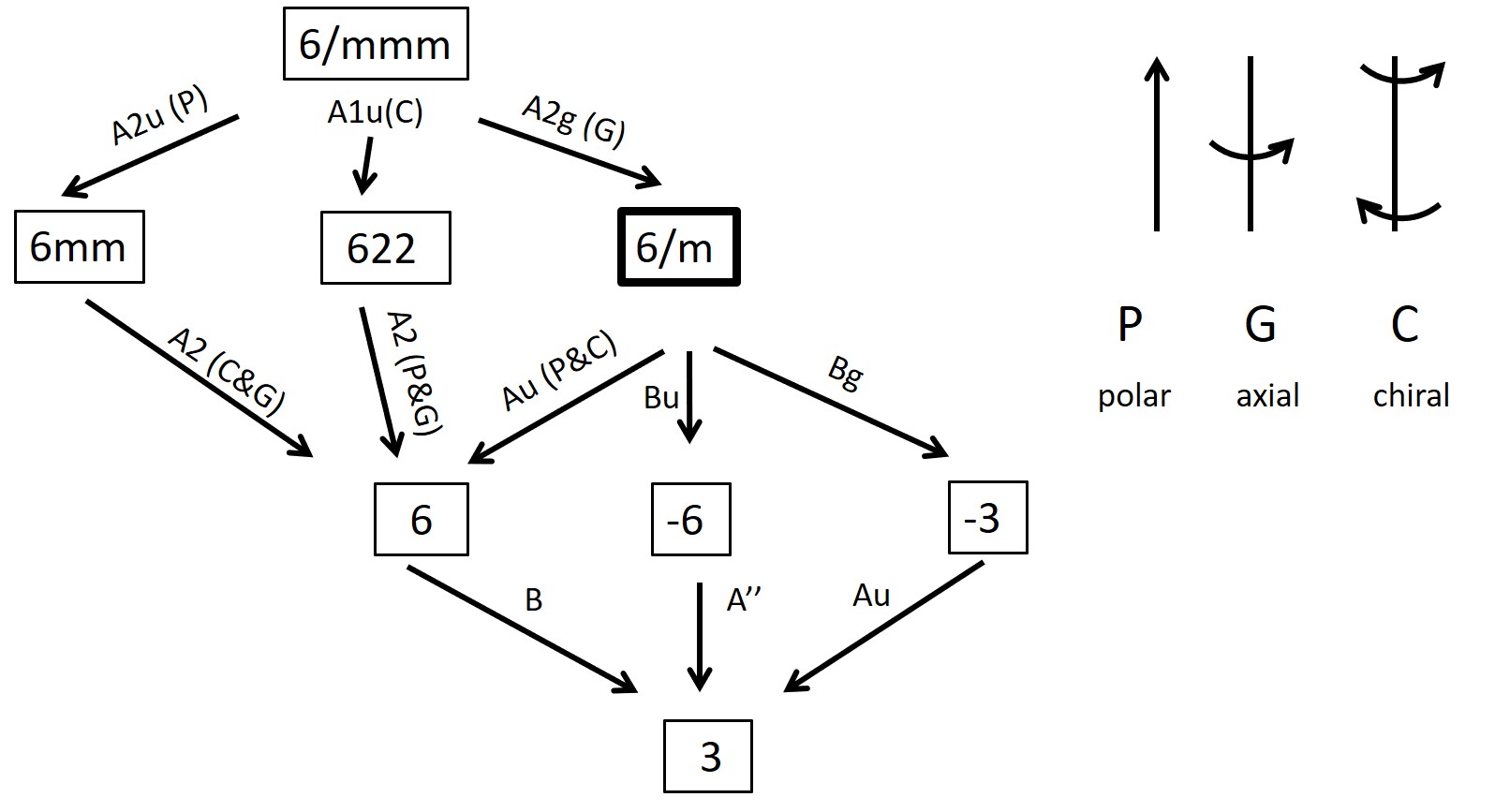}
\caption{Graph of point group symmetry reductions corresponding to Fig.\,3.
}
\label{ctvrty}
\end{figure}

The  $P6_3, P\bar{3}, P\bar{6}, P2_1, P2_1/m, P2_1/b, P6_3cm$ and $Pnma$ spacegroups, covering practically all other reported apatite structures, can be also seen as symmetry descents of this common aristotype\cite{White05}. Diagram showing subordination of all hexagonal subgroups of this aristotype is shown in Fig.\,3. All symmetry reductions displayed in Fig.\,3  correspond to the point group symmetry lowering. Therefore, these symmetry reductions  can be classified as ferroic species\cite{Hlinka2016}, and there is one-to-one correspondence to the point-group reductions shown in Fig.\,4. Moreover, the minimal symmetry breaking marked by arrows in Fig.\,3 are induced by $\Gamma$-point irreducible representations indicated in Fig.\,4. The $P6_3/mcm > P6_3/m$ transition to the phase of Fig.\,1 can be induced by $A_{2g}$ lattice mode. It transforms as an axial vector parallel to the hexagonal axis, which can be represented for example as an electric toroidal moment $G$ of a set of dipoles arranged around an axis similarly as the tetrahedra in Fig.\,2. 
Therefore, the  $P6_3/mcm > P6_3/m$ transition is a clear example of the so-called ferroaxial phase transition\cite{Hlinka2016}.

Alternatively, the symmetry of the aristotype can be 
 lowered by a polar $A_{2u}$ lattice mode to the polar group $P6_3cm$, or by a chiral $A_{1u}$ lattice mode to the chiral group $P6_322$. These order parameters transform as polar vector or chiral bidirector\cite{Hlinka2014}, so that the corresponding phase transitions can be considered as ferroelectric and ferrochiral ones\cite{Erb18,Erb20}. While the ferroelectric  $P6_3cm$ phase has been found in several Ba$_{10}$(ReO$_5$)$_6$X$_2$ apatites\cite{White05, Abrahams88}, we are not aware of any $P6_322$ apatites. 
 
 The nature of the symmetry breaking by polar vector (P), axial vector (G) and chiral bidirector (C) is intuitively obvious from their pictograms (see Fig.\,4). The trilinear coupling term among parallel P, G and C symmetry order parameters transforms as a fully symmetric representation of the Curie group of their common axis\cite{Hlinka2014}. It is therefore always allowed in free energy expansion of uniaxial crystals, what implies that whenever any two of the P, G, C order parameters are frozen, the third one of the triplet should be nonzero as well. Here such triple condensation to the polar-axial-chiral subgroup is realized in the $P6_3$ (173) crystal structure. 

 As a side note, let us mention that the Landau free energy expansion for $P6_322$ apatite structure allows a linear term in polarization gradients, transforming as chiral bidirector along $z$-axis. Such term is a  direct electric analogue of the chiral Dzialoshinski-Moria interaction responsible for stability of the bulk magnetic Bloch skyrmion textures in  continuous Bogdanov-Yablonskii theory as explained in Ref.\,\onlinecite{Erb20}. 
 Consequently, the $P6_3$ ferroelectric subgroup of $P6_322$,  with a spontaneous polarization along $z$-axis, could potentially host electric skyrmions,  mathematically analogous to those of the original Bogdanov-Yablonskii theory\cite{Erb20}, even if the high symmetry phase is not accessible experimentally. In that respect, one promising electric skyrmion candidate could be the calcium sulfoapatite\cite{Suitch85, Abrahams90}.

\begin{figure}[h]
\includegraphics[width=.95\columnwidth]{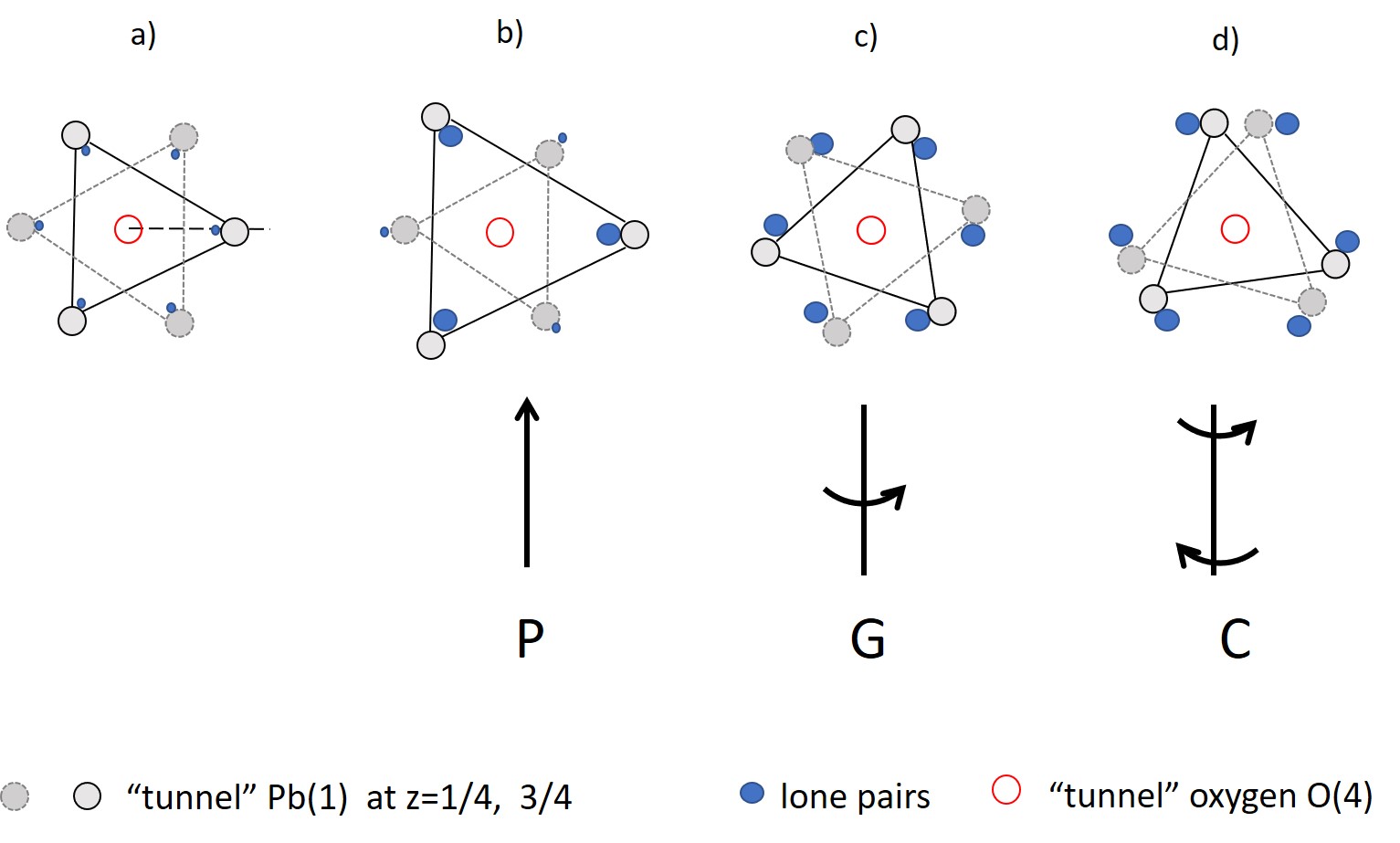}
\caption{Symmetry adapted distortion modes of Pb$^{\rm T}$ cation coordination around the "tunnel" axis.
}
\label{treti}
\end{figure}

\section{Lone pair charge density wave}
Lets come back to the structure of lead apatite.
Several authors have drawn the attention to stereoactivity of the lead  lone pairs in apatites\cite{Peet17, Cametti22} as well as in relation to the copper-doping induced shifts of the tunnel lead cations in the Pb$_9$Cu(PO$_4$)$_6$O  apatite\cite{Lee233a,Lee236a, Griffin23}. 
In general,  the lone-pair off-centering is associated with slight displacement corresponding lead ion,
so that the charge-density wave should be coupled to phonon wave, whatever is the primary cause of the final state.
It can be expected that Pb$^{\rm T}$(1) lone pairs in apatites with large tunnels or strongly electronegative fillers would be preferentially oriented towards the tunnel axis, keeping the hexagonal symmetry of the tunnel, as schematically charted in Fig.\,5a. 
However, since the three Pb(1) cations are located in the mirror symmetry plane at $z=1/4$ and the other three Pb(1) cations at $z=3/4$, there is also enough space for tangential lone pair off-centering along the circumference of the tunnel. 
For example, clockwise tangential off-centering of lone pairs on both layers would lead to a  toroidal (G) symmetry charge density wave, while the clockwise off-centering at one layer combined with the anti-clockwise one at the other layer would define a chiral (C) symmetry charge density wave (see Fig.\,5c,d).
Clearly, the charge density wave with symmetry of a $z$-axis axial vector would not break the symmetry of the $P6_3/m$ structure since it already contains the toroidal distortion. 

At the same time, inspection of  the analysis  the diffraction experiment given in the original report\cite{Lee233a,Lee236a} suggests that the lead displacements between layers are in antiphase, but rather in the radial direction, suggesting the arrangement like in Fig.\,5b, which has symmetry of polar vector parallel to the $z$-axis (see Fig.\,3c of Ref.\onlinecite{Lee233a}). 
On the other hand, the recent theoretical paper\cite{Griffin23} proposed that the copper-doping triggered charge density should be chiral. 
In fact,  we can conclude that the either polar or chiral distortion is enough to decrease the symmetry at least to 
the $P6_3$ phase. In other words, the charge density wave with a symmetry of polar vector would induce also a chiral distortion while the charge density wave with a symmetry of a chiral bidirector would induce also a polar distortion (of $P6_3/m$). 
Therefore, the metal-insulator phase transition induced by copper doping seems to be also ferrochiral and ferroaxial one.

Note that some ferroic properties are obviously related to the possibility of switching domain states by means of macroscopic external stimuli like stress or electric fields, and here, obviously, electric switching among the domains will be difficult to achieve due to the screening of the electric field in the metallic state and first order coupling to the strain is fobidden by symmetry.

\section{Other sources of symmetry breaking}

Another source of symmetry breaking could be  the ordering of copper ions substituted at  Pb$^{\rm F}$ positions, or the ordering of the extra oxygen in the "tunnel" sector. As far as we know, the available experiments on the Pb$_9$Cu(PO$_4$)$_6$O apatite does not show evidence for such ordering, although the correlation between positions of the copper and the extra oxygen ions is probably needed for the anticipated physics phenomena.

Basic  plausible structures with a long-range positional order are obviously those selected in the trial configurations explored in density functional calculations. For example, occupation of only one of the four equivalent positions of the 4e Wyckoff site of the extra oxygen decreases the symmetry to 
$P3$. Same effect has the replacement of one particular Pb$^{\rm F}$ ion by the copper. Resulting configurations correspond to the orientational domains of the corresponding point group symmetry lowering, and therefore also differ by signs of certain components macroscopic tensors\cite{Hlinka2016}. 

\section{Conclusion}

In summary, we have shown that the available experimental and theoretical studies indicate  that the average crystal structure of copper-substitution to lead phosphate apatite is decreased to $P6_3$, implying that the metallic phase is simultaneously polar and chiral and has ferroic properties of macroscopic crystallographic symmetry breaking species $6/m > 6$. The noncentrosymmetric character of the high-temperature metallic phase might be important for fine features of the eventual mechanism of the so far hypothetic superconductivity of the low-temperature phase.

\section*{Acknowledgements}

This work was supported by the Czech Science Foundation (project no. 19-28594X).


\end{document}